\documentstyle[prl,aps,multicol]{revtex}
\begin{document}
\draft
\preprint{}

\renewcommand{\narrowtext}{\begin{multicols}{2}}
\renewcommand{\widetext}{\end{multicols}}
\newcommand{\be}{\begin{equation}}
\newcommand{\ee}{\end{equation}}
\title{Ultra-high energy cosmic rays without GZK cutoff}

\author{V. Berezinsky$^1$, M. Kachelrie{\ss}$^1$, and A.Vilenkin$^2$}

\address{$^1$INFN, Laboratori Nazionali del Gran Sasso,
             I--67010 Assergi (AQ), Italy}

\address{$^2$Institute of Cosmology, Department of Physics and Astronomy,
             Tufts University, Medford, MA 02155, USA}

\maketitle

\begin{abstract}
We study the decays of ultraheavy ($m_X \geq 10^{13}~GeV$) and quasistable 
(lifetime $\tau_X$ much larger than the age of the Universe $t_0$)
particles as 
the source of Ultra High Energy Cosmic Rays (UHE CR). These particles are 
assumed to constitute a tiny fraction $\xi_X$ of CDM in the Universe, with   
$\xi_X$ being the same in the halo of our Galaxy and in the intergalactic 
space. The elementary-particle and cosmological scenarios for these 
particles are briefly outlined. The UHE CR fluxes produced at the decays of X-
particles are calculated. The dominant contribution is given by fluxes 
of photons and nucleons from the halo of our Galaxy and thus they do not 
exibit the GZK cutoff. 
The extragalactic components of UHE CR are 
suppressed by the smaller extragalactic density of X-particles and hence 
the cascade limit is relaxed. We discuss the spectrum of produced Extensive 
Air Showers (EAS) and a signal from Virgo cluster as signatures of
this model.
\end{abstract}


\narrowtext
The observations of Ultra-High Energy Cosmic Rays (UHE CR) reveal the 
presence of a new, isotropic component at energies $E\geq 1\cdot 10^{10}$~GeV 
(for a review see Ref. \cite{nagano}). This component is thought to have an 
extragalactic origin since the galactic magnetic field cannot isotropize 
the particles of such energies produced by astrophysical sources in
the Galaxy. On the other hand, the observation of particles of
the highest energies, especially of the two events with energies $2 - 3 \cdot 
10^{11}$~GeV  \cite{highest}, contradicts the 
GZK cutoff \cite{GZK} at $E\sim 3\cdot 10^{10}$~GeV,
which is the signature of extragalactic UHE CR. All known extragalactic 
sources of UHE CR, such as AGN \cite{birm}, topological defects \cite{td} or 
the Local Supercluster \cite{bg}, result in a well pronounced GZK cutoff,
although in some cases the cutoff energy is shifted
closer to $1\cdot 10^{11}$~GeV \cite{bg}.
 UHE neutrinos \cite{bz} could give a spectrum 
without cutoff, but the neutrino fluxes and the neutrino-nucleon 
cross-section are not large enough to render the neutrino a realistic
candidate for the UHE CR events. 
 
In this {\em Letter}, we propose a scenario in which
the UHE CR spectrum has no GZK cutoff and is nearly  isotropic. 
Our main assumption is that Cold Dark Matter 
(CDM) has a small admixture of long-lived supermassive $X$-particles.
Since, apart from very small scales, fluctuations grow identically
in all components of CDM, the fraction of $X$-particles, $\xi_X$, is 
expected to be the same in all structures. In particular, $\xi_X$ is
the same in the halo of our Galaxy and in the
extragalactic space. Thus the halo density of $X$-particles is enhanced in 
comparison with the extragalactic density.
The decays 
of these particles produce UHE CR, whose flux is dominated by the 
halo component, and therefore has no GZK cutoff. Moreover,
the potentially dangerous cascade radiation \cite{ps}
is suppressed. Long-lived massive relic particles were already 
discussed in the literature as a source of high energy neutrino radiation
\cite{llmp}. However, in our case the particles must be much heavier 
($m_X \sim 10^{13} - 10^{16}$~GeV).

The plan of our paper is as follows. First, we take an phenomenological
approach and treat the density $n_X$ of $X$-particles 
and their lifetime $\tau_X$ as free parameters fixed only by the
requirement that the observed UHE CR flux is reproduced.
We calculate the fluxes of nucleons, photons and neutrinos,
considering the production of cascade radiation, positrons, antiprotons and 
radio fluxes as constraints. We then discuss how the required 
properties of $X$-particles can be realized.

The decays of $X$-particles result in the production of nucleons with a 
spectrum $W_N(m_X,x)$, where $m_X$ is the mass of the X-particle 
and $x=E/m_X$. The flux of nucleons $(p,\bar{p},n,\bar{n})$ from the halo and 
extragalactic space can be calculated as
\be
I_N^{i}(E)={1\over{4\pi}}{n_X^i\over{\tau_X}}R_i{1\over{m_X}}W_N(m_X,x),
\label{eq:nhalo}
\ee
where index $i$ runs through $h$ (halo) and $ex$ (extragalactic), 
$R_i$ is the size of the halo $R_h$, or the attenuation length of 
UHE protons due to their collisions with microwave photons,      
$\lambda_p(E)$, for the halo case and extragalactic case, respectively. We 
shall assume $m_Xn_X^h=\xi_X\rho_{\rm CDM}^h$ and 
$m_Xn_X^{\rm ex}=\xi_X\Omega_{\rm CDM}\rho_{\rm cr}$, 
where $\xi_X$ describes the fraction 
of $X$-particles in CDM, $\Omega_{\rm CDM}$ is the CDM 
density in units of the critical density $\rho_{\rm cr}$. 
We shall use the following values for these parameters: a large 
DM halo with $R_h=100$~kpc (a smaller halo with $R_h=50$~kpc is possible, 
too), $\Omega_{CDM}h^2=0.2$, 
the mass of $X$-particle in the range 
$10^{13}~{\rm GeV}<m_X<10^{16}$~GeV, 
the fraction of $X$-particles 
$\xi_X\ll 1$ and $\tau_X \gg t_0$, where $t_0$ is the age of the Universe.
The two last parameters are convolved in the flux calculations in a single
parameter $r_X=\xi_X t_0/\tau_X$. Following \cite{BMV},
we shall use the QCD fragmentation function in MLLA approximation
(see \cite{MLLA})
\be
W_N(m_X,x)=\frac{K_N}{x}\exp \left( -\frac{\ln^2 x/x_m}{2\sigma^2} \right),
\label{frag}
\ee
where
$$
2\sigma^2=\frac{1}{6} \left( \ln \frac{m_X}{\Lambda} \right) ^{3/2},
$$
$x=E/m_X$, $x_m=(\Lambda/m_X)^{1/2}$ and $\Lambda=0.234$~GeV. The 
normalization constant $K_N$ is found from energy conservation as
$$
K_N \int_0^1 dx \exp \left( -\frac{\ln^2 x/x_m}{2\sigma^2} \right)=f_N ,
$$
where $f_N$  is the fraction of energy transferred to nucleons. 
Using $Z^0$-decay as a guide, 
we assume $f_N \approx 0.05 f_{\pi}$, where $f_{\pi}$ is 
the corresponding fraction for pions (LEP gives 0.027 for $p\bar{p}$ only).
For the attenuation length of UHE protons due to their interactions with
microwave photons, we use the values given in the book \cite{BBGDP}. 

The high energy photon flux is produced mainly due to decays of neutral 
pions and can be calculated for the halo case as
\be
I^h_{\gamma}(E)=\frac{1}{4\pi}\frac{n_X}{\tau_X}R_h N_{\gamma}(E),
\label{gflux}
\ee
where  
$N_{\gamma}(E)$ is the number of photons with energy $E$ produced per
decay of one $X$-particle. The latter is given by 
\be
N_{\gamma}(E)=\frac{2K_{\pi^0}}{m_X}\int_{E/m_X}^1 \frac{dx}{x^2}
\exp \left( -\frac{\ln^2 x/x_m}{2\sigma^2} \right) .
\label{gnumber}
\ee
The normalization constant $K_{\pi^0}$ is again found from the condition that 
neutral pions take away the fraction $f_{\pi}/3$ of the total energy $m_X$.

For the calculation of the extragalactic gamma-ray flux, it is enough to 
replace the size of the halo, $R_h$, by the absorption length of a photon, 
$\lambda_{\gamma}(E)$. The main photon absorption process is $e^+e^-$ pair 
production on background radiation and, at $E>1\cdot 10^{10}$~GeV,
on the radio background. The neutrino flux calculation is similar.

Before discussing the obtained results, we consider 
various astrophysical constraints.

The most stringent constraint comes from electromagnetic 
cascade radiation,
which is initiated by high-energy photons and electrons from pion decays 
and is developing due to interaction with low energy background photons.
The relevant calculations were performed in Ref. \cite{ps}. In our case 
this constraint is weaker, because the low-energy extragalactic
nucleon flux 
is $\sim 4$ times 
smaller than that one from the Galactic halo (see Fig.~1). Thus 
the cascade radiation is suppressed by the same factor. 

The relevant parameter which characterizes the flux of cascade radiation is 
the total energy density of cascade radiation $\omega_{\rm cas}$. 
The observation of the low-energy diffuse gamma-ray flux results in
the limit $\omega_{\rm cas} < 1\cdot 10^{-5} - 1\cdot 10^{-6}$~eV/cm$^3$ 
\cite{ps}. In our case, the 
cascade energy density calculated by integration over cosmological epochs
(with the dominant contribution given by the present epoch $z=0$) yields
\be
\omega_{\rm cas}=\frac{1}{5}r_X\frac{\Omega_{CDM}\rho_{cr}}{H_0t_0}=
6.3\cdot10^2 r_X f_{\pi}~{\rm eV/cm}^3.
\label{cas}
\ee

To fit the UHE CR observational data by nucleons from halo, 
we need $r_X=5\cdot 10^{-11}$. Thus the cascade energy density is 
$\omega_{\rm cas}=3.2\cdot 10^{-8} f_{\pi}$~eV/cm$^3$, well below the
observational bound. 

The other constraints come from the observed fluxes of positrons and 
antiprotons in our Galaxy and from the isotropic component of the radio flux. 
We performed detailed calculations which will be published elsewhere. 
In all cases the abovementioned constraints are satisfied and they are 
weaker than that due to cascade gamma-radiation.

Now we address the elementary-particle and cosmological aspects
of a supermassive, long-living particle. Can the relic density 
of superheavy $X$-particles be as high as required in our calculations?
And can this particle have a lifetime comparable or larger than the age
of the Universe?
 
Let us assume that $X$ is a neutral fermion which belongs 
to a representation of the $SU(2)\times U(1)$ group.  We assume also
that the stability of $X$-particles is protected by a discrete
symmetry 
which is respected by all interactions except quantum
gravity through wormhole effects. In other words, our particle is very
similar to a very heavy neutralino with a conserved quantum number, $R'$, 
being the direct analogue of $R$-parity (see \cite{BJV} and the
references therein).
Thus, one can assume that the decay of $X$-particles occurs due to dimension 5
operators, inversely proportional to the Planck mass $m_{\rm Pl}$ and 
additionally suppressed by a factor $\exp(-S)$, where $S$ is the 
action of a wormhole which absorbs one $R'$-charge. 
As an example one can consider a term
\be
{\cal L} \sim \frac{1}{m_{Pl}} \bar{\Psi}\nu \phi\phi \exp(-S),
\label{eq:d5}
\ee
where $\Psi$ describes X-particle, and $\phi$ is a $SU(2)$ scalar with vacuum 
expectation value $v_{EW}=250$~GeV.
After spontaneous symmetry breaking the term (\ref{eq:d5}) results in 
the mixing of $X$-particle and neutrino, and  
the lifetime due to $X \to \nu +q + \bar{q}$ , {\it e.g.}, is given by
\be
\tau_X \sim \frac{192(2\pi)^3}{(G_Fv_{EW}^2)^2}\frac{m_{\rm Pl}^2}{m_X^3}
e^{2S},
\label{eq:ltime}
\ee
where $G_F$ is the Fermi constant.  The lifetime $\tau_X >t_0$ for 
$X$-particle with $m_X \geq 10^{13}$~GeV needs $S>44$. 
This value is within the 
range of the allowed values as discussed in Ref. \cite{KLLS}.

Let us now turn to the cosmological production of $X$-particles with 
$m_X \geq 10^{13}$~GeV. Several mechanisms can be considered, including 
thermal production at the reheating stage, production through the decay of 
inflaton field at the end of the "pre-heating"
period following inflation, and through the decay of hybrid topological 
defects, such as monopoles connected by strings or walls bounded by
strings.  

For the thermal production, temperatures comparable to $m_X$ are needed. 
In the case of a heavy decaying gravitino,
the reheating temperature $T_R$ (which is the highest temperature 
relevant for our problem)            
is severely limited to value below $10^8- 10^{10}$~GeV, depending 
on the gravitino mass (see Ref. \cite{ellis} and references therein).  
On the other hand, 
in models with dynamically broken supersymmetry, the lightest 
supersymmetric particle is the gravitino. Gravitinos with mass 
$m_{3/2} \leq 1$~keV  interact relatively strongly with the thermal bath,
thus decoupling relatively late, and can be the CDM particle \cite{grav}. 
In this scenario all phenomenological
constraints on $T_R$ (including the decay of the second 
lightest supersymmetric particle) disappear and one can assume
$T_R \sim 10^{11} - 10^{12}$~GeV. In this 
range of temperatures, $X$-particles are not in thermal equilibrium.
If $T_R < m_X$, the density  $n_X$ of $X$-particles produced during the 
reheating phase at time $t_R$ due to $a+\bar{a} \to X+\bar{X}$ is easily 
estimated as
\be
n_X(t_R) \sim N_a n_a^2 \exp\left(-\frac{2m_X}{T_R}\right)\sigma_X t_R,
\label{eq:dens}
\ee 
where $N_a$ is the number of flavors which participate in the production of 
X-particles, $n_a$ is the density of $a$-particles and $\sigma_X$ is 
the production cross-section. The density of $X$-particles at the
present epoch can be found by the standard procedure of calculating
the ratio $n_X/s$, where 
$s$ is the entropy density. Then for $m_X = 1\cdot 10^{13}$~GeV
and $\xi_X$ in the wide range of values $10^{-8} - 10^{-4}$, the required
reheating temperature is $T_R \sim 3\cdot 10^{11}$~GeV.

In the second scenario mentioned above, non-equilibrium inflaton decay,
$X$-particles are usually overproduced and a second period of 
inflation is needed 
to suppress their density.

Finally, $X$-particles could be produced by the decay
of hybrid topological defects, {i.e.\/} monopoles connected by strings or
walls bounded by strings.  For example, strings of energy scale
$\eta_s\gtrsim m_X$ could be formed at a phase transition at or near the
end of inflation.  At a second phase transition with symmetry-breaking
scale $\eta_w <m_X$ each string gets attached to a domain
wall. The wall tension pulls
the strings together and eventually leads to a breakup of the network.
The resulting pieces of wall bounded by string lose
their energy by gravitational radiation and by particle production.
$X$-particles are produced whenever strings cross one another and also
in the decay of the pieces which fragmented down to the size
comparable to the string thickness.  The $X$-particle density produced
in this way depends on the details of the fragmentation process, but
rough estimates suggest that the required values of $n_X/s$ can be
obtained for a wide range of string and wall parameters.

Let us now discuss the obtained results.
The fluxes shown in Fig.~1 are obtained for $R_h=100$~kpc, 
$m_X=1\cdot 10^{13}$~GeV and
$r_X=\xi_X t_0/\tau_X=5\cdot10^{-11}$. This ratio $r_X$ allows very small 
$\xi_X$ and $\tau_X > t_0$. The fluxes 
near the maximum energy $E_{\rm max}=5\cdot 10^{12}$~GeV  were only roughly
estimated (dotted lines on the graph). 

It is easy to verify that the extragalactic nucleon flux at $E \leq 
3\cdot 10^{9}$~GeV is suppressed by a factor $\sim 4$ and by a much larger 
factor at higher energies due to nucleon energy losses. The flux of 
extragalactic photons is suppressed even stronger, because the attenuation 
length for photons (due to absorption on radio-radiation) is much smaller
than for nucleons (see Ref. \cite{PB}). This flux is not shown in the graph.  
The flux of  high energy gamma-radiation from the halo is by a factor $7$ 
higher than that of nucleons and the neutrino flux, given in the Fig.1 as 
the sum of the dominant halo component and subdominant extragalactic one,
is twice higher than the gamma-ray flux.

The spectrum of the observed EAS is formed due to fluxes of gamma-rays and 
nucleons. The gamma-ray contribution to this spectrum is rather complicated.
In contrast to low energies, the photon-induced showers at 
$E>10^9$~GeV have the low-energy muon component as abundant as that 
for nucleon-induced showers \cite{AK}. However, the  
shower production by the photons is suppressed by the 
LPM effect \cite{LPM}
and by absorption in geomagnetic field (for recent calculations and 
discussion see \cite{ps,Kasa} and references therein). These effects are 
energy dependent. The LPM effect 
starts at $10^9 - 10^{10}$~GeV and it almost
fully suppresses the production of "normal" EAS at 
$E_{\gamma} \geq 1\cdot 10^{12}$~GeV, when maximum EAS reaches the see level 
practically for all zenith angles \cite{ps}. The calculation of the 
spectrum of EAS is outside the scope of this paper and the normalization
of the halo nucleon spectrum by observational data at $E \sim 2\cdot 
10^{11}$~GeV in Fig.~1 has an illustrative character. The general
tendency of greater 
suppression of photon-induced showers with increase of energy
might improve the agreement between calculated and observed spectra. 

We wish to note that the excess of the gamma-ray flux over the nucleon
flux  from the halo is an unavoidable feature of our model. It follows
from the more effective production of pions 
than nucleons in the QCD cascades from the decay of $X$-particle. 


Although $X$-particles with necessary properties can be produced by a
variety of mechanisms, it should be noted that their lifetime and
spatial density had to be fine-tuned
to the desired values with the help of exponential factors.

The signature of our model  might be the signal from the Virgo 
cluster. The virial mass of the Virgo cluster is
$M_{\rm Virgo} \sim 1\cdot 10^{15} M_{\odot}$ and the distance to it 
$R= 20$~Mpc. If UHE protons (and antiprotons) propagate rectilinearly from 
this source
(which could be the case for $E_p \sim 10^{11} - 10^{12}$~GeV), their 
flux is given by
\be
F_{p,\bar{p}}^{\rm Virgo}= r_X \frac{M_{\rm Virgo}}{t_0 R^2 m_X^2}W_N(m_X,x).
\ee    
The ratio of this flux to the diffuse flux from the half hemisphere is
$6.4\cdot 10^{-3}$. This signature becomes less pronounced at smaller 
energies, when protons can be strongly deflected by intergalactic magnetic 
fields.
 

When our work was in progress, we learned that  a similar
idea was put forward by V. A. Kuzmin and V. A. Rubakov \cite{dubna}. 
The main difference is that we take into account the radiation from
the galactic halo, which is the main issue of our work, 
while the  authors above limited their
consideration to the 
extragalactic component. We are grateful to V.A. Kuzmin and V.A. Rubakov 
for interesting discussions.
M.K. was supported by a Feodor-Lynen scholarship of the Alexander
von Humboldt-Stiftung.

\widetext



\end{document}